\documentclass[12pt]{iopart}

\usepackage{iopams}  
\usepackage{hyperref}
\usepackage{float}
\usepackage{xspace}
\usepackage{graphicx}
\usepackage{dcolumn}
\usepackage{bm}
\usepackage{color}
\usepackage{soul}
\usepackage{multirow}
\usepackage{url}
\usepackage{xcolor}

\begin{document}

\title[]{Novel wide spectrum light absorber heterostructures based on hBN/In(Ga)Te}

\author{A.~\v{S}olaji\'{c} and J.~Pe\v{s}i\'{c}}
\address{Center for Solid State Physics and New Materials, Institute of Physics Belgrade, University of Belgrade, Pregrevica 118, 11080 Belgrade, Serbia}
\ead{solajic@ipb.ac.rs}

\vspace{10pt}
\begin{indented}
\item[]April 2022
\end{indented}

\begin{abstract}
Two dimensional group III monochalcogenides have recently attracted quite attention for their wide spectrum of optical and electric properties, being promising candidates for optoelectronic and novel electrical applications, however in their pristine form they are very sensitive and vulnerable to oxygen in air. Here we present two newly designed vdW heterostructures based on hBN (hexagonal boron nitride) and GaTe or InTe monolayer. Using density functional theory we investigate electronic and optical properties of those structures. Their moderate band gap and an good absorption coefficient makes them excellent for absorbers in wide spectrum, covering all from part of IR to far UV spectrum, with particularly good absorption of UV light. The hBN layer which can be beneficial for protection of sensitive GaTe and InTe does not only preserve their optical properties but also enhances it. Moreover, we confirm that type of stacking does not affect any relevant properties, as all three stacking types are very similar in total energy and bandstructure is almost the same for every one. This is especially important for easier experimental realization.
\end{abstract}

%
%
%
%
%

\section{Introduction}

In recent years, enormous attention given to the exploration and researching of two-dimensional materials has started a whole new era in materials science and countless possibilities for novel devices emerged. After graphene, many similar van der Waals (vdW) layered materials were researched \cite{liu2019recent} - hexagonal boron nitride \cite{zhang2017two,song2010large}, silicene \cite{kara2012review,zhao2016rise}, germanene \cite{acun2015germanene,derivaz2015continuous}, transition metal dichalcogenides (TMD's) \cite{fu20212d,li2011mos2,splendiani2010emerging,li2012bulk,zhao2013evolution,li2013mechanical}, MXenes \cite{anasori20172d,pang2019applications}. That also opened a door for even more innovations and research of 2D vdW heterostructures, which can be attractive for countless applications in nanoelectronic and optoelectronic devices \cite{withers2015light,geim2013van,novoselov20162d,liu2016van}.

Single layer group III monochalcogenides have been extensively researched recently \cite{ren2019first,demirci2017structural,yang2019recent,chen2020high}. As layered quasi 2D structures with weak interlayer binding forces, they are suitable for mechanical exfoliation to few layers or even monolayer. They have a wide spectra of great electronic and optical properties, making them very desirable candidates for further research and applications. For example, the most popular among them, InSe exhibits high electron mobility \cite{bandurin2017high} due to which can be excellent for applications in field effect transistors \cite{jiang2019stable} but also possess good optical properties and can be used for photodetectors from UV to the near IR region \cite{tamalampudi2014high}.
Many vdW heterostructures based on those compounds are also being studied - for example group III monochalcogenides with graphene as an tunable Schottky barrier \cite{li2020tunable,li2019schottky,pham2019modulation,pham2018layered,gao2019graphene}, and recently heterostructures based on InSe/hBN showed as good absorbers of the visible and UV part of spectrum.  

Recently, two members of group III monochalcogenides were explored and theoretically proposed as new 2D structures - monolayers of InTe and GaTe. Both are indirect band gap semiconductors with band gaps of 1.29  eV and 1.75 eV, respectively. In addition to good electrical and optical properties \cite{touski2020electrical,liao2020electronic,li2019first,ariapour2019spin,vi2020modulation}, those materials also excel in elastic properties, being able to sustain considerable values of  both  tensile and compressive strain \cite{touski2020electrical,jalilian2017electronic,vi2020modulation}, which showed as very effective and convenient way to precisely tune electrical and optical properties of 2D materials \cite{zhang2018strain,xiong2020strain,postorino2020strain}. Suitable for various applications in novel electronic and optoelectronic devices, research of heterostructures based on those materials is very attractive, expecting to achieve some new effects or to tune and enhance desired properties of the 2D structures alone. However, pristine monochalcogenides, especially in thin films or single layer, are very sensitive and vulnerable to oxygen in air - many studies reveal that monolayers are oxidized fast, almost instantly, after exposure to the air \cite{Rahaman2017,guo2020oxidation, guo2017defects,afaneh2020large}. The issue of their challenging stability can be overcome by passivization with adequate material which can ensure the safe encapsulation of monochalcogenides as well a good mechanical protection. One of materials shown as particularly good for this purpose is hexagonal boron nitride (hBN) \cite{bandurin2017high}. Experimental studies confirmed it as effective for protection and passivization of few layer InSe and GaSe, while their electronic and optical properties are preserved or even enhanced \cite{arora2019effective}. 

In the next sections, we present newly designed heterostructures, hBN/GaTe and hBN/InTe, based on single layer of GaTe or InTe and layer of hBN. Using the density functional theory, we explore their electronic and optical properties and analyse the influence of hBN on GaTe and InTe.

\section{Theoretical Methods}
Results are obtained using density functional theory (DFT) implemented in Quantum Espresso (QE) software package \cite{QE-2009}, based on plane waves and pseudopotentials. Generalized gradient approximation (GGA) and PAW pseudopotentials were used in all calculations. After convergence tests, the energy cutoff for the wavefunction and the charge density were set to 44 Ry and 364 Ry for hBN/InTe heterostructure, and 60 Ry and 480 Ry for hBN/GaTe heterostructure. In order to simulate 2D structure, a vacuum of 20 $\mathrm{\mathring{A}}$ was added along the z-direction to avoid interactions between the layers. As the GGA functionals do not take into consideration long range forces as the van der Waals force, Grimme-D2 correction was included to obtain more accurate lattice constants and forces. Optical properties were calculated using epsilon.x code in QE software, based on the random phase approximation (RPA). 

\section{Results and discussion}

\begin{figure}
	\centering
	\includegraphics[width=0.7\linewidth]{"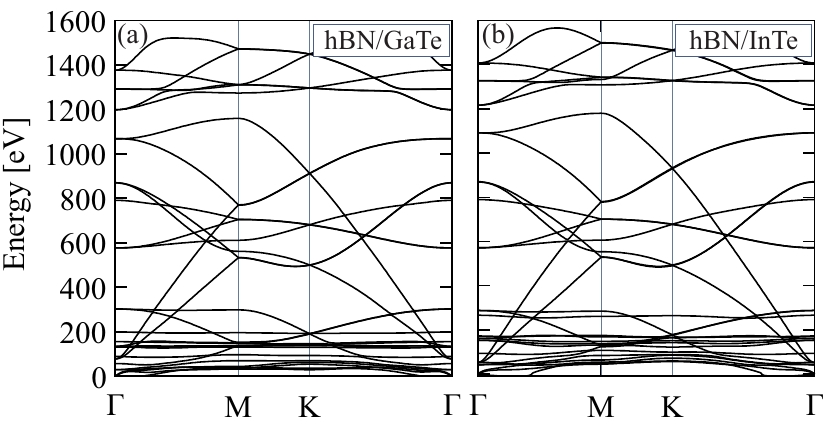"}
	\caption{Phonon dispersions of (a) hBN/InTe and (b) hBN/GaTe.}
	\label{fig:phdisp}
\end{figure}

Both GaTe (InTe) and hBN have hexagonal lattice with $D_{3h}^1$ symmetry. Lattice constants obtained after the geometric optimizations of $a=4.37\mathrm{\mathring{A}}$ for InTe, $a=4.05\mathrm{\mathring{A}}$ for GaTe and $2.51\mathrm{\mathring{A}}$ for hBN are in agreement with previous reports \cite{demirci2017structural,vi2020modulation,liao2020electronic,wickramaratne2018monolayer}. The unit cells of In(Ga)Te/hBN heterostructures contain of one layer of InTe(1x1) or GaTe(1x1) and one layer of hBN($\sqrt{3}\times\sqrt{3}$). Resulting unit cells are hexagonal lattice cells with $C_{3}^1$ symmetry, with lattice constants of $a=4.34\mathrm{\mathring{A}}$ for hBN/InTe and $a=4.31\mathrm{\mathring{A}}$ for hBN/GaTe. In terms of lattice matching, both heterostructures are promising and can be modelled with supercell like here - induced strains in those heterostructures are 6.1\% on GaTe and 0.8\% on hBN in hBN/GaTe, whereas 0.8\% of strain on InTe and 0.3\% on hBN in hBN/InTe makes it especially promising for experimental realization. Phonon dispersion for both heterostructures is also calculated and presented in Figure \ref{fig:phdisp} in order to confirm  the structural stability. We do not observe imaginary frequencies, except the small kinks near the Gamma point with very small negative values which are often emerging in calculations of phonons in 2D materials, being a numerical issue and not the real instabilities.

\begin{figure}
	\centering
	\includegraphics[width=0.7\linewidth]{"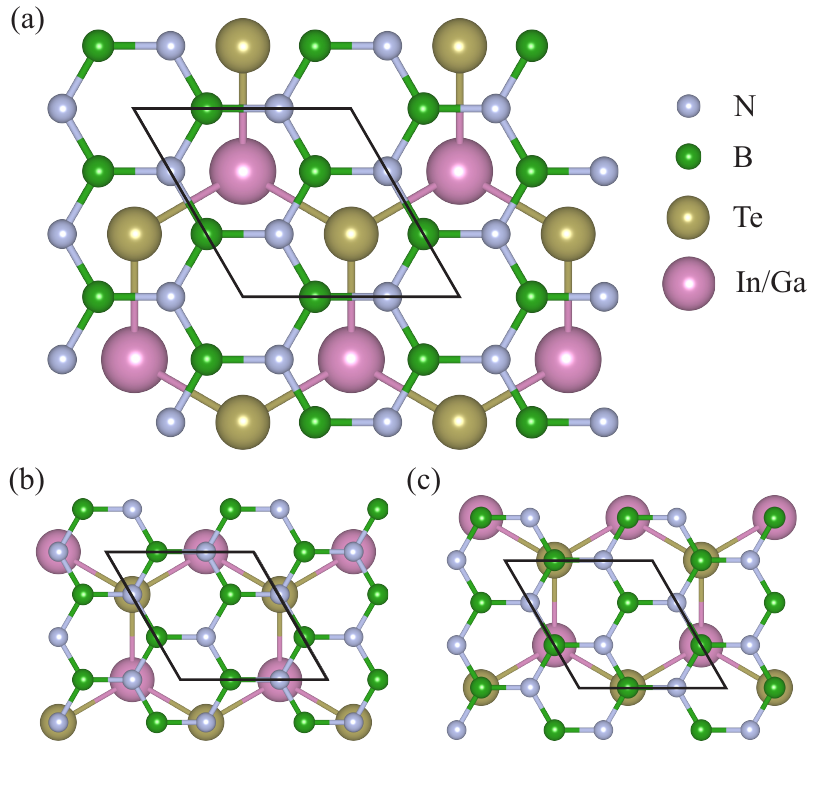"}
	\caption{Top view of three possible stacking types, (a) H-top, (b) N-top and (c) B-top.}
	\label{fig:struktura}
\end{figure}

Three possible stacking types are presented in Fig \ref{fig:struktura}, the H-top (In/Ga atom being in the centre of hBN hexagon), B-top (In/Ga atom above the B atom of hBN) and N-top (In/Ga atom above the N atom of hBN). We investigated all three types of structure in order to determine the favourable stacking, as well whether the properties are affected by type of stacking, e.g bandstructure. After geometry optimization with both atom positions and lattice constants relaxed with forces up to $10^{-6} \mathrm{Ry/\mathring{A}}$, we conclude that both heterostructures favour the H-top type of stacking with respect to the total energy (which is lower for ~10 meV for a H-top stacking type) and their binding energies. Lattice parameters for all three stacking types for both structures are summarized in Table \ref{tab:parameters}.

\begin{table}
	\centering
	\caption{Lattice parameters and distance between hBN and In(Ga)Te ($d$) for all three possible stacking types of hBN/InTe and hBN/GaTe.}
	\label{tab:parameters}
	\vspace{5pt}
	\begin{tabular}{|c|c|c|c|}
		\hline
		& \multicolumn{3}{c|}{hBN/InTe} \\
		\hline
		& H-top & N-top & B-top \\
		\hline
		$a [\mathrm{\mathring{A}}]$ & 4.3362 & 4.337 & 4.337 \\
		\hline
		$d [\mathrm{\mathring{A}}]$ & 3.429 & 3.523 & 3.479 \\
		\hline
		& \multicolumn{3}{c|}{hBN/GaTe} \\
		\hline
		& H-top & N-top & B-top \\
		\hline
		$a [\mathrm{\mathring{A}}]$ & 4.309 & 4.309 & 4.311 \\
		\hline
		$d [\mathrm{\mathring{A}}]$ & 3.451 & 3.516 & 3.503 \\
		\hline
	\end{tabular}
\end{table}

\begin{figure}
	\centering
	\includegraphics[width=0.7\linewidth]{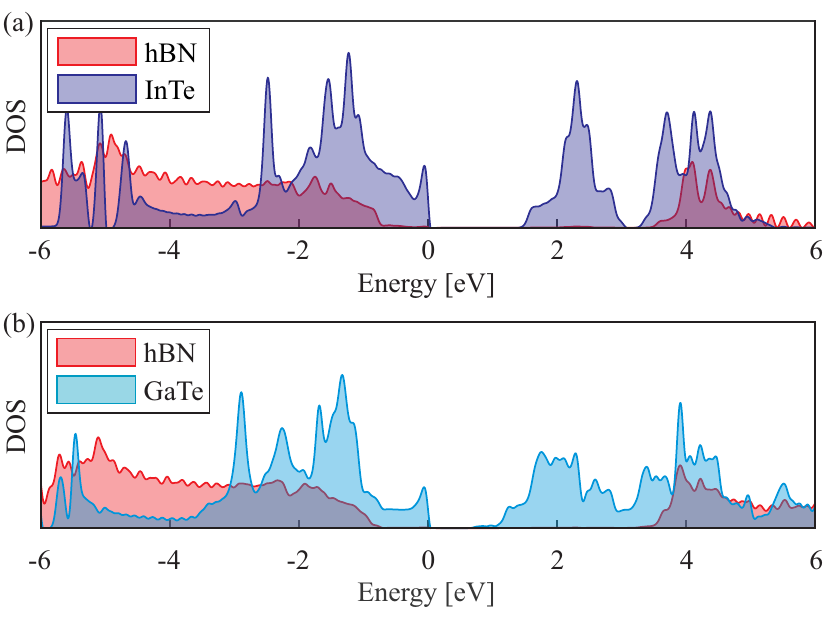}
	\caption{Projected density of states for (a)hBN/InTe and (b) hBN/GaTe heterostructure. The contribution from hBN is shown in red lines and red shaded area, while dark blue and turquoise lines and area represent contribution from InTe and GaTe, respectively.}
	\label{fig:dos}
\end{figure}

\begin{figure}
	\centering
	\includegraphics[width=0.7\linewidth]{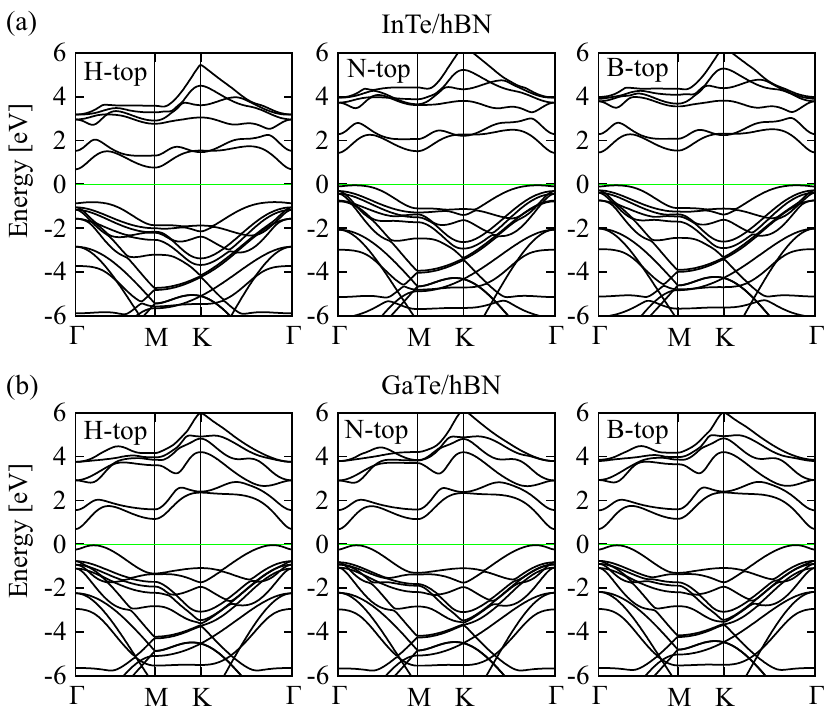}
	\caption{Bandstructures of different types of stackings for (a) hBN/InTe and (b) hBN/GaTe heterostructure.}
	\label{fig:bands-stacking}
\end{figure}

From our calculations, both InTe and GaTe monolayers have an indirect bandgap of $E_g=1.38$ eV and $E_g=1.75$ eV respectively, while hBN has a large direct bandgap of 4.63 eV. These results are in agreement with previous theoretical results obtained using the PBE functional \cite{touski2020electrical,liao2020electronic,ren2019first,vi2020modulation}. As the PBE functional underestimates the band gap in semiconductors, hybrid functionals such is HSE must be used in order to obtain an accurate electronic properties. Reports on similar structures show that employing the HSE functional does not change the bandstructure qualitatively, most significant difference comes from shifted bands above the Fermi level and thus an enlargement of band gap. Large difference in bandgap of InTe(GaTe) and hBN as well their alignment of bands make both systems a type-I heterojunctions. This can be also confirmed from projected density of states presented in Figure \ref{fig:dos}.
As in similar heterostructures \cite{shen2020two}, the bandstructure does not change with the stacking almost at all - zones near the Fermi level are almost identical, the bandgap does not change and only minor differences can be observed e.g. slightly changed position of some zones below the Fermi level and far above the Fermi level. Bandstructure plots for all three stacking types are shown in Figure \ref{fig:bands-stacking}. In further discussion we will focus on the H-top stacking in both hBN/InTe and hBN/GaTe heterostructures.

\begin{figure}
	\centering
	\includegraphics[width=0.7\linewidth]{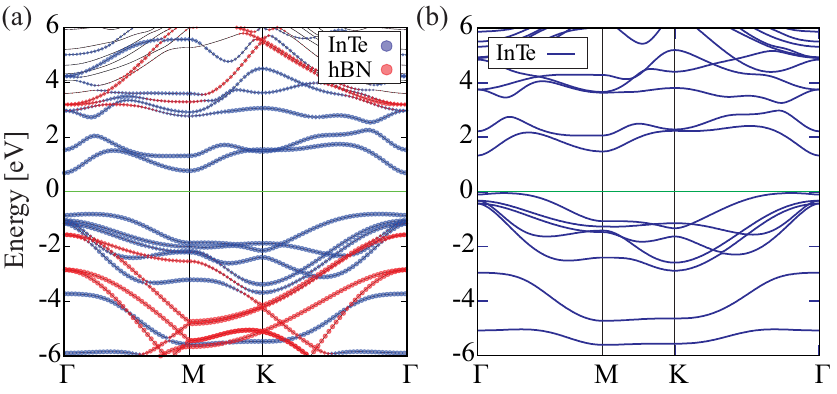}
	\caption{Bandstructure of (a) hBN/InTe heterostructure and (b) InTe monolayer. The size of circles represent the amount of InTe contribution (blue circles) and hBN contribution (red circles).}
	\label{fig:bands-in}
\end{figure}

\begin{figure}
	\centering
	\includegraphics[width=0.7\linewidth]{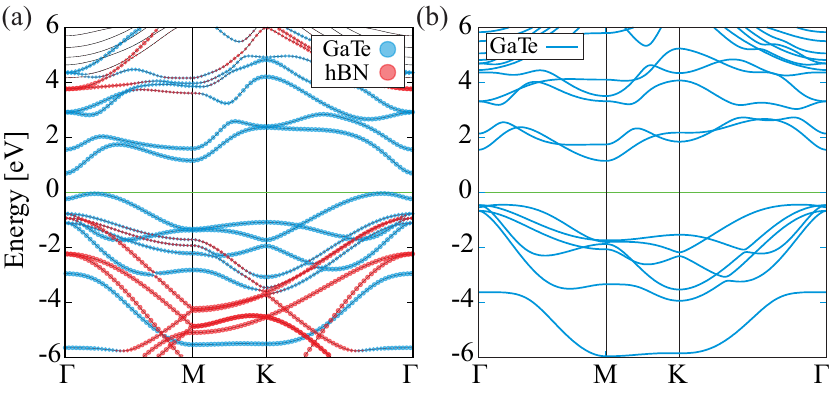}
	\caption{Bandstructure of (a) hBN/GaTe heterostructure and (b) GaTe monolayer. The size of circles represent the amount of GaTe contribution (light blue circles) and hBN contribution (red circles).}
	\label{fig:bands-ga}
\end{figure}

Projected bandstructures of hBN/InTe and hBN/GaTe are presented in Figures \ref{fig:bands-in} and \ref{fig:bands-ga}, alongside their pristine monolayer compounds, InTe and GaTe. Fermi level is set to zero and shown in green line. Both GaTe and InTe pristine monolayers are indirect band gap semiconductors with valence band maximum (VBM) close to the $\Gamma$ point and conduction band minimum (CBM) at the M point for GaTe and $\Gamma$ and M points for InTe. After stacking into heterostructure, the band gap is slightly changed from original GaTe and InTe structures. Conduction band minimums are also now in the $\Gamma$ point in both heterostructures. In the hBN/InTe band gap is slightly enlarged, from 1.38 eV in pure InTe to 1.54 eV in the heterostructure. Upon forming the hBN/InTe heterostructure, except slightly smaller band gap, there are no important differences in the band structure. States around the Fermi level are almost completely formed by InTe, while the hBN contributions are observed below -2 eV and above 3 eV, similar as in pristine hBN.

More changes can be observed in the second structure - gap of single layer GaTe is 1.59 eV, but after the heterostructure is formed, new states emerge near the $\Gamma$ point right below the Fermi level, distributed mainly from GaTe, reducing the band gap to just 0.79 eV.

The reduction of band gap along with interfacial states can be useful for applications in optoelectronics so we proceeded to analyse the optical properties of these heterostructures. We calculate the absorption coefficients, expecting of both structures to have good absorption properties. The complex dielectric function $\epsilon(\omega)+\epsilon_R(\omega)+i\epsilon_I(\omega)$ is obtained from calculations in the RPA framework. From these results, we can obtain the absorption coefficient $\alpha(\omega)$ as the following:
\begin{equation}
	\alpha(\omega) = \sqrt{2}\frac{\omega}{c}\sqrt{\sqrt{\epsilon_R^2(\omega)+\epsilon_I^2(\omega)}-\epsilon_I(\omega)}
\end{equation}

\begin{figure}
	\centering
	\includegraphics[width=0.7\linewidth]{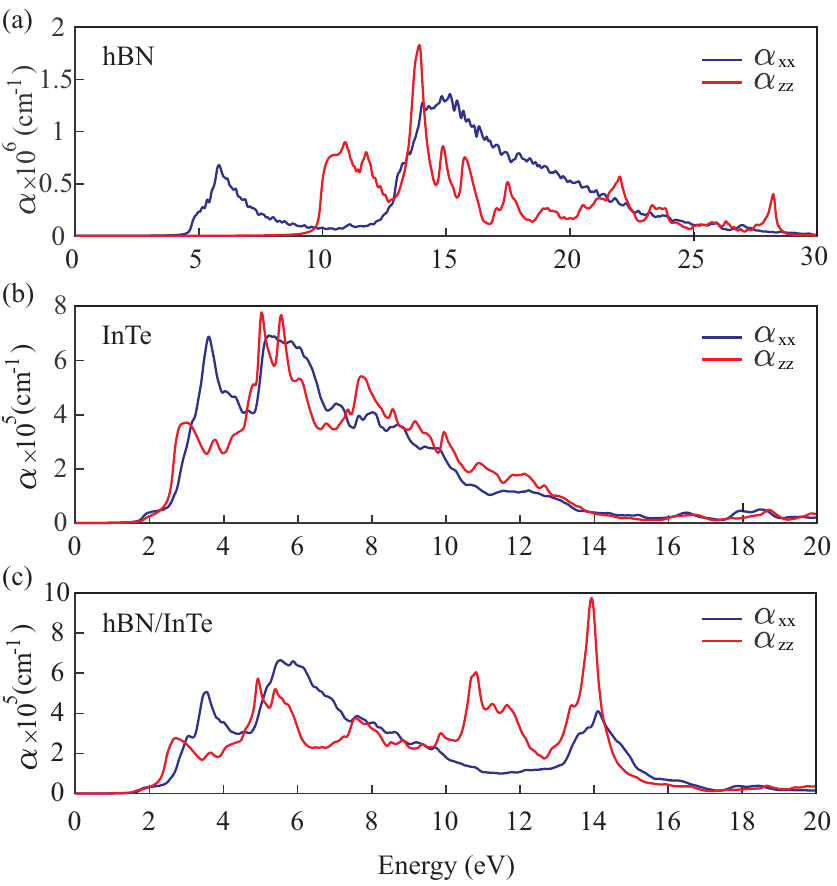}
	\caption{Absorption coefficient of (a) hBN (b) InTe (c) hBN/InTe heterostructure for in-plane ($\alpha_{xx}$) and out-of-plane ($\alpha_{zz}$) polarizations.}
	\label{fig:abs-in}
\end{figure}

\begin{figure}
	\centering
	\includegraphics[width=0.7\linewidth]{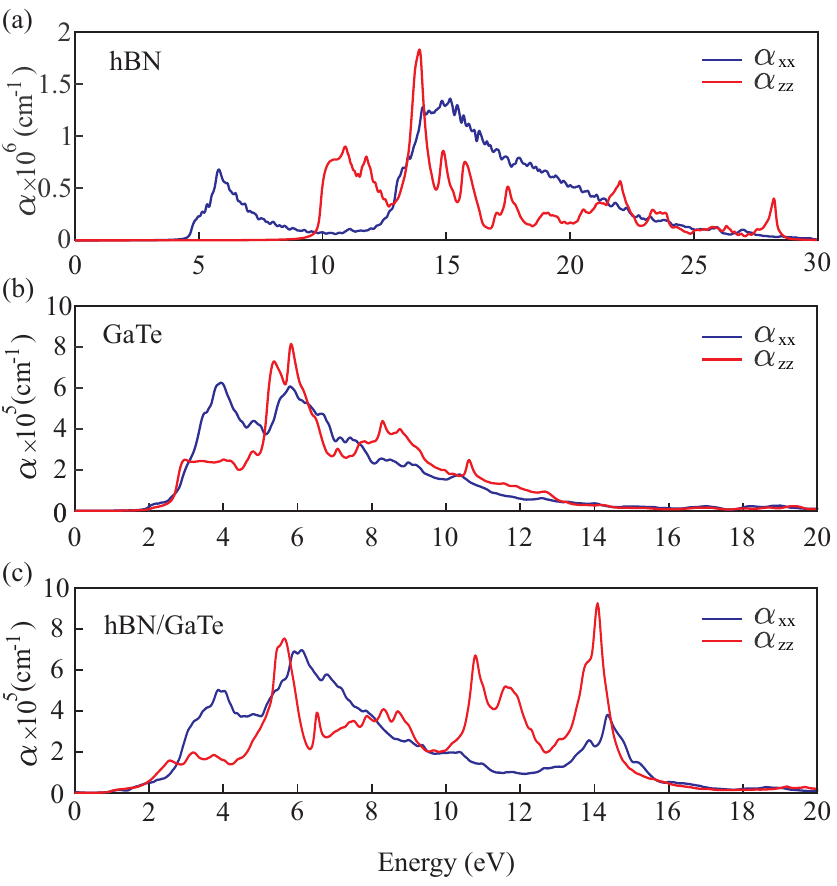}
	\caption{Absorption coefficient of (a) hBN (b) GaTe (c) hBN/GaTe heterostructure for in-plane ($\alpha_{xx}$) and out-of-plane ($\alpha_{zz}$) polarizations.}
	\label{fig:abs-ga}
\end{figure}

Results are shown in Figure \ref{fig:abs-in} for hBN/InTe and \ref{fig:abs-ga} for hBN/GaTe heterostructures, both plotted alongside their pristine monolayer compounds. Both structures have very good absorption properties, being able to absorb visible, near and far UV spectrum - the absorption of hBN/GaTe also extends to the IR spectrum. As the PBE functional underestimates band gap, some shift in energy would be noticed compared to HSE calculations. The results though would not be significantly changed, only the reduction in capabilities of absorbing the IR and red part of the visible light is expected.

The highest peaks in absorption function are near 6 eV, primarily originating from the InTe/GaTe, at 11 eV and at 14 eV as a contribution from hBN layer, with peaks up to $8\times 10^5 \, \mathrm{cm^{-1}}$. In addition, in energy range from 3 eV to 15 eV, both heterostructures have exceptional values of absorption of $2-8 \times 10^5 \, \mathrm{cm^{-1}}$, giving them a huge potential for light absorbers in near and far UV spectrum. This makes both heterostructures an excellent candidates for light absorption-relevant photoelectric applications. We also calculated the reflectance for our heterostructures. The real and imaginary part of dielectric function are related with a complex index of refraction $n^*(\omega)=n(\omega)+i\kappa(\omega)$ by:
\begin{equation}
	\epsilon_R = 2n\kappa, \, \epsilon_I = n^2 - \kappa^2.
\end{equation}
The reflectance is now given by
\begin{equation}
	\frac{(n-1)^2 + \kappa^2}{(n+1)^2+\kappa^2}.
\end{equation}

\begin{figure}
	\centering
	\includegraphics[width=0.7\linewidth]{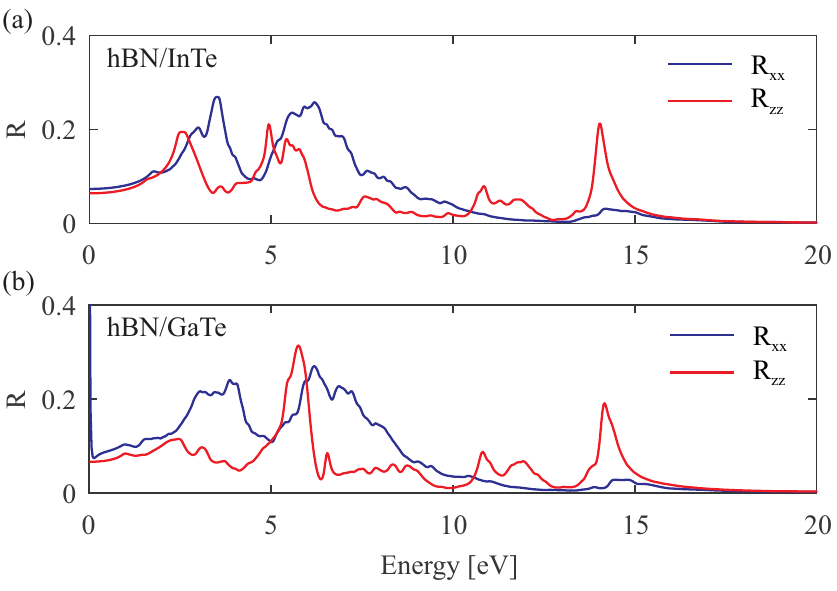}
	\caption{Reflectance of (a) hBN/InTe heterostructure and (b) hBN/GaTe, for in-plane ($\alpha_{xx}$) and out-of-plane ($\alpha_{zz}$) polarizations.}
	\label{fig:reflectance}
\end{figure}

Figure \ref{fig:reflectance} presents calculated reflectance of hBN/InTe and hBN/GaTe heterostructures. Reflectance of both materials are no larger than 30\% for any energy, and are especially low in the low-energy region with ~10\%, confirming that only a small amount of light is being reflected at any incidence angle. 

\section{Conclusions}
We report the study of two new heterostructures based on hexagonal boron nitride and InTe/GaTe, by employing first principle calculations. We confirm the structural stability of those systems and show that stacking type does not affect any of the relevant properties. We investigate their electronic and optical properties, showing the benefit of forming a heterostructures of InTe/GaTe with a hBN layer. Formation of heterostructure slightly reduces the bandgap in comparison with pristine monolayer InTe and GaTe, which also enhances absorption in the low-energy region of spectrum. The electronic and optical properties of those heterostructures reveal them as an excellent candidates for various optoelectronic devices with great capabilities of absorption from visible light to far UV part of spectrum, being exceptionally good for absorbing the UV light. From the previous studies, the hBN layer is also proven to be beneficial for mechanical protection of sensitive and vulnerable single layers of monochalcogenides like InTe and GaTe, while as we showed,  electronic and optical properties are not only preserved but even enhanced.
As both InTe (GaTe) and hBN can withstand a moderate strain, electronic and optical properties can be tuned by applying the strain on the heterostructure or by including electric field, giving these materials a huge value for further research and applications.

\section{Acknowledgements}
The authors acknowledge funding provided by the Institute of Physics Belgrade through the grant by the Ministry of Education, Science and Technological Development of the Republic of Serbia. DFT calculations were performed using computational resources at Johannes
Kepler University (Linz, Austria).


\section*{References}
\bibliography{References}   
\bibliographystyle{iopart-num.bst}  

\end{document}